\documentstyle[psfig,amssymb]{mn}

\title[The driving mechanisms for spiral structure]
{Near--infrared constraints on the 
driving mechanisms for spiral structure.}

\author[Seigar et al.] 
{Marc S. Seigar$^{1, 2}$, Nicole E. Chorney$^{1, 3}$ \& Phil A. James$^4$\\  
$^1$Joint Astronomy Centre, University Park, 660 North A'ohoku Place,
Hilo, HI 96720, USA\\ 
$^2${\tt email: m.seigar@jach.hawaii.edu}\\
$^3$Department of Physics \& Astronomy, University of Victoria, PO Box 3055 STN CSC, Victoria BC, V8W 3P6, Canada\\
$^4$Astrophysics Research Institute, Liverpool John Moores University, Twelve Quays House, Egerton Wharf, Birkenhead CH41 1LD, UK
}

\begin{document} 
\maketitle

\begin{abstract}

We have imaged a sample of 17 inclined spiral galaxies with measured H$\alpha$
rotation curves in the {\em K} band, in order to determine the morphology of
the old stellar population which dominates the mass in the disc. The {\em K}
band images of the galaxies have been used to determine the radial extent of
Grand--Design spiral structure and compare this with the turnover radius in
their rotation curves, where the rotation curve transforms from solid--body
rotation to differential rotation. Thus, if the arms do not extend past this
radius, the winding problem is solved. We find in all 17 cases, the arms extend
past this radius, with the radius of grand--design spiral structure being
a factor of 1.3--9.6 times larger than the rotation curve turnover radius. 
Of these galaxies, 4 have nearby neighbours and central bars
and a further 7 have a central bar. These bars or near neighbours may be the
cause of the driving of the spiral potential in the discs of these galaxies. Of
the remaining 6 galaxies, 5 show some evidence for a bar or oval
distortion in their {\em K} band images. The remaining galaxy (UGC 14) shows 
no evidence for a central bar and has no nearby neighbours.

Finally we also find that the spiral structure of these galaxies in the
near--infrared is extremely regular, although some range in the regularity
of spiral structure is found. To quantify this range in spiral structure, 
we introduce the dust--penetrated
arm class, which is analogous to the dust--penetrated classification scheme
of Block \& Puerari (1999) and based on the optical arm class of Elmegreen
\& Elmegreen (1982).
\end{abstract}
  
\begin{keywords}
galaxies: fundamental parameters -- galaxies: spiral
-- infrared: galaxies
\end{keywords}

\section{INTRODUCTION}

In the last decade, near--infrared photometry has changed our view of spiral
galaxies, revealing different arm structures from optical imaging (e.g.
Block \& Wainscoat 1991; Block et al. 1994; Thornley 1996). Specifically, the
near--infrared gives a much truer view of the underlying mass distribution,
and shows that Grand Design spiral structure is more common than optical
studies would suggest (Thornley 1996). Apart from the clearer view of the
strength and extent of arms, the near--infrared also frequently reveals
central bar structures which were invisible in the optical, and again
gives a more quantitative view of the dynamical importance of bars (Seigar
\& James 1998b; Eskridge et al. 2000; Seigar 2002). Motivated by this, we
feel that it is timely to revisit the suggestion of Kormendy \& Norman (1979)
that spiral structure can be explained without the need for density wave
theories, posing the much more stringent test set by near--infrared data.

Several explanations have been proposed for the existence of Grand--Design
spiral structure in disc galaxies. The most widespread and commonly accepted
explanation is the idea of a quasi--stationary density wave 
(Lin \& Shu 1964, 1966)
and further developments such as the modal theory of spiral structure
(Bertin et al. 1989a, b). However, a proposal made over twenty years ago by
Kormendy \& Norman (1979), but often overlooked, 
suggested that if Grand--Design spiral structure in
disc galaxies does not extend past the radius at which the rotation curve
turns over, then the arms can indeed be material arms and they
will not wind up under solid--body rotation, thus overcoming the winding 
problem. Kormendy \& Norman (1979) also 
proposed that if any galaxies did exhibit Grand--Design spiral structure
outside this radius, then they must do so under the influence of a 
central bar--like
potential, or under the tidal influence of a nearby neighbouring galaxy.
Indeed out of the 54 galaxies observed by Kormendy \& Norman (1979), they
found only 2 galaxies which could not be explained by their hypothesis.

Observational support for this view has been provided by Elmegreen \&
Elmegreen (1987) who observed that, generally, grand--design spiral galaxies
are found in richer environments than flocculent spirals (i.e. grand--design
spiral structures are more likely to have undergone a tidal interaction with 
another galaxy in their history).

While this postulation does not necessarily rule out density wave theories
(e.g. Lin \& Shu 1964, 1966; Bertin et al. 1989a, b), it introduces an 
alternative
mechanism for exciting spiral modes, i.e. either bar or tidally driven
modes. Indeed the original hypothesis of a quasi-stationary density wave
(Lin \& Shu 1964) proposed that a density wave is
the result of a natural instability of a cold, self--gravitating, stellar disc.
However, modal theories (Bertin et al. 1989a, b; Bertin 1991, 1993, 1996;
Bertin \& Lin 1996) of spiral structure do not
rule out the possibility of bar or tidally driven modes.

The ideas of tidally driven spiral structure in disc galaxies, have been 
studied in simulations as early as those of 
Toomre \& Toomre (1972) who performed
an N--body simulation of M51 and its interacting galaxy. More recent
N--body simulations of this system include those of Salo \& Laurikainen 
(2000a, b).
While the
spiral structure in M51 probably pre--existed the encounter with its
companion (Zaritsky, Rix \& Rieke 1993), 
M51 shows a very strong spiral structure, which may
have been enhanced by this interaction. Evidence was presented by
Elmegreen, Elmegreen \& Seiden (1989) which showed that the arm--interarm
contrast in M51 is significantly larger than in two other galaxies (M81
and M100) which do not possess companions. This suggest that the large
arm--interarm contrast in M51 may be a result of its interaction with
NGC 5195. 

Sundelius et al. (1987) used the code of Miller (1976a, b, 1978) to produce 
a model of tidal spiral arms in two--component galaxies. Their model consisted
of a self--gravitating disc of 60,000 particles and this disc was stabilised
by a ``halo'' component. The disc consisted of two components, the cold 
component, representing the stars in spiral arms (e.g. OB stars, gas clouds 
etc) and the hot component, representing old stars. The different components
have different initial velocity dispersions. The disc was perturbed by a point
mass passing by in a plane parabolic orbit. This leads to the formation of a 
2--armed spiral pattern in the cold component of the disc as a result of tidal
triggering by the companion. The pattern survives for 6 rotations, until the
velocity dispersion of the cold component is about half that of the hot 
component. The pitch angles of the arms are seen to evolve from an Sc--like
appearence to an Sa--like appearance. Donner \& Thomasson (1994) used N--body 
simulations to study non--linear evolution of tidally--induced spiral patterns
in galaxies. They found that the patterns live with a nearly constant amplitude
for about 5 to 6 rotation periods, and are then regenerated after disappearing
momentarily.

It is
therefore expected that galaxies that have undergone an interaction with
another galaxy, may have enhanced spiral structure, and their spiral arms
will appear stronger than those galaxies without nearby neighbours (although
this may only be a transient phenomena). Indeed,
in an earlier paper (Seigar \& James 1998b) we showed evidence to suggest
that this might be the case. Out of 45 galaxies observed in the near--infrared,
12 had near neighbours. Of these galaxies 4 out of 7 galaxies with arm
strength (defined as equivalent angle, EA, see Seigar \& James 1998a and 
section 3.1 for
a definition of EA) greater than 35$^{\circ}$ had near neighbours. This showed
that there is some indication here of a relation between arm strength and 
tidal interactions, although the difference is not statistically significant.
However, this result was strengthened by the fact that galaxies with near
neighbours had an enhanced $m=2$ mode, this being the dominant mode in 7 out
of the 12 galaxies with near neighbours. 

Bar driving of spiral structure was modelled as early as Sanders \& Huntley
(1976). They constructed a uniform gas disc with no self--gravity. Initially,
the gas was in circular oribits. After the introduction of a potential due to
a rigidly rotating bar, the gas settled into a steady state exhibiting a 
strong trailing spiral pattern. In this simulation there was no need for
a density wave. However, there are at least two problems with bar driving of
spiral structure. Firstly, bar driving seems unable to produce tightly wound
spiral patterns. Also, the spiral pattern is seen in the old disc stars as 
well as young stars and gas, yet the Sanders \& Huntley (1976) simulation
predict a spiral pattern for the gas only. For example, the model of Sanders
(1977) could not produce a spiral pattern for a simulated disc of test 
particles. More recently the N--body simulations
of Sellwood \& Sparke (1988) and Sanders \& Tubbs (1980) have shown that
bar forcing should not extend far outside the region where the bar is strong.

In our earlier paper (Seigar \& James 1998b) we investigated the relationship
between bar strength and arm strength. We showed that there was no evidence of
a correlation between these two quantities. If spiral structure was driven
by a bar potential, one might expect these two quantities to be well
correlated. As a result, we concluded that bar driving was not an important
factor in the formation of spiral structure in disc galaxies.

In this paper we introduce a sample of galaxies observed in the near--infrared,
with a view to testing the ideas of Kormendy \& Norman (1979). These galaxies
all have measured rotation curves. In section \ref{sec:obs} we introduce the
sample; in section \ref{sec:res} we summarize the results of a test the hypothesis
made by Kormendy \& Norman (1979). We treat their hypothesis as having
three distinct parts, and test them accordingly:
\begin{itemize}
\item a comparison of the radius of extent of spiral structure with the turnover
radius in the rotation curve.
\item a test of whether spiral structure is induced by tidal interactions with neighbouring galaxies.
\item a test of bar driven spiral structure.
\end{itemize}
Finally in section \ref{sec:conc} we summarize our conclusions.

\section{Observations}
\label{sec:obs}

\begin{figure*}
\caption{Greyscale {\em K} band images of the galaxies in the sample 
classified as non--barred.}
\end{figure*}

\begin{table*}
\caption{The sample of galaxies observed in the K--band with their Hubble classifications and near neighbours where appropriate. Column 1 lists the names of the galaxies; Column 2 lists the Hubble types of the galaxies from de Vaucouleurs et al. (1991); Column 3 lists the name of the companion galaxy when present; Column 4 lists the redshift difference in kms$^{-1}$ between the object galaxy and the companion galaxy; Column 5 gives $r_{gd}/r_{to}$, the ratio of the radius of Grand Design spiral structure to the rotation curve turnover radius; Column 6 lists the arm EA for the galaxies; Column 7 lists the bar EA for those galaxies with bars; Column 8 lists the dust--penetrated arm class of each galaxy; and Column 9 has further comments on the spiral structure of some of the galaxies.}
\begin{center}
\begin{tabular}{l l l l l l l c l}
\hline
Galaxy		& Hubble 	& Near 			& Redshift			& $r_{gd}/r_{to}$	& Arm		& Bar		& Dust--		& Comments	\\
		& type		& neighbour		& difference			&			& EA		& EA		& penetrated		&		\\
		&		&			& (kms$^{-1}$)			&			& (degrees)	& (degrees)	& arm class		&		\\
\hline
ESO 515--G3	& SB(rs)c	&	--		&		& 2.7		& 23$\pm$4	& 21$\pm$3	& 1			&	\\
ESO 543--G12	& SB(s)ab	&	--		&		& 4.2		& 5$\pm$2	& 40$\pm$5	& 2			&	\\
ESO 555--G8	& SB(s)c	& IC 438		& 9.00		& 1.9		& 4$\pm$2	& 15$\pm$3	& 2			&	\\
ESO 574--G33	& SB(rs)bc	&	--		&		& 2.3		& 19$\pm$3	& 60$\pm$6	& 1			&	\\
ESO 576--G12	& SA(rs)c	&	--		&		& 1.3		& 5$\pm$1	& --		& 1			&	\\
ESO 576--G51	& SB(s)bc	& IRAS 13218--1929	& 97.00		& 6.1		& 20$\pm$3	& 46$\pm$4	& 1			&	\\
ESO 583--G2	& SB(rs)bc	&	--		&		& 3.8		& 11$\pm$2	& 45$\pm$3	& 1			&	\\
ESO 583--G7	& SB(rs)c	&	--		&		& 3.7		& 37$\pm$4	& 58$\pm$5	& 3			& 	\\
ESO 602--G25	& SA(r)b	&	--		&		& 9.3		& 30$\pm$4	& --		& 1			&	\\
ESO 606--G11	& SB(rs)bc	&	--		&		& 1.9		& 5$\pm$2	& 17$\pm$2	& 1			&	\\
IC 1330		& Sc? sp	&	--		&		& 5.9		& 25$\pm$4	& --		& 1			& 3--armed pattern	\\
NGC 2584	& SB(s)bc?	& FAIRALL 0273		& 78.00		& 3.1		& 20$\pm$3	& 31$\pm$3	& 1			&	\\
NGC 2722	& SA(rs)bc	&	--		&		& 2.1		& 20$\pm$6	& --		& 3			& Several arm fragments		\\
NGC 7677	& SAB(r)bc	& VV 619		& 77.00		& 1.51		& 17$\pm$3	& 7$\pm$1	& 2			& 	\\
UGC 14		& Sc+		&	--		&		& 9.6		& 10$\pm$1	& --		& 1			& Several knots	\\
UGC 210		& Sb		&	--		&		& 2.7		& 5$\pm$1	& --		& 1			&	\\
UGC 12383	& SABb		&	--		&		& 2.3		& 15$\pm$2	& 9$\pm$1	& 2			&	\\
\hline
\end{tabular}
\end{center}
\end{table*}

\begin{figure*}
\includegraphics{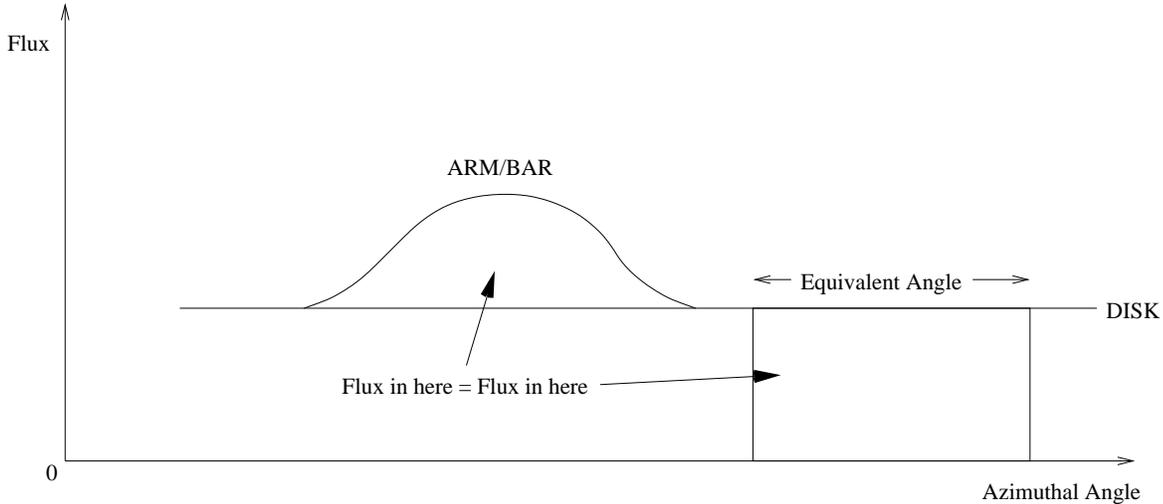}
\vspace*{6.5cm}
\caption{Diagrammatic representation of equivalent angle.}
\end{figure*}

We have observed 17 galaxies on the UK Infrared Telescope (UKIRT) 
in 2 runs between 1st -- 4th August 2001 and 11th -- 14th March 2002 
in the {\em K} band (2.2 $\mu$m) using
the UKIRT Fast Track Imager (UFTI). UFTI is a 1--2.5 $\mu$m camera
consisting of 1024$\times$1024 pixels of 0.09 arcsec each. Thus its field
of view is $\sim$90 arcsec. The galaxies were therefore selected such that
they could fit on the UFTI field. The galaxies were selected from the
Mathewson et al. (1992) sample. These galaxies all have measured H$\alpha$
rotation curves (unlike the sample of Seigar \& James 1998a, for which no
rotation curve data was available, because the galaxies were too face--on).
They were selected such that their declination, $\delta >
-20^{\circ}$, and the ratio of their major--to--minor axes, $a/b<3.0$, so that
the galaxies are not too edge on and spiral structure can still be traced.
The galaxies were not selected on the appearance of their spiral arm structure.

In the first run from 1st--4th August 2001, galaxies were observed down
to the K=21.5 magnitude isophote with $S/N=3$ per arcsec$^2$. 
In the latter run, from
12th--14th March 2002, galaxis were observed down to the K=22.1 magnitude
isophote with $S/N$=3 per arcsec$^2$.

The justification for using the near--infrared {\em K} band is two--fold.
Firstly, the {\em K} band is relatively free from dust extinction when
compared to optical wavebands. Also, the {\em K} band traces the old
stellar population, and thus the dominant mass distribution in galaxy discs.
This is important when studying the underlying structure in disc
galaxies.

\section{Results}
\label{sec:res}

It should first be noted that the {\em K} band image of each galaxy reveals 
a regular spiral arm structure, formally known as Grand Design. This was not
part of the selection process, and therefore shows that, although galaxies
can appear flocculent in the optical, a much more regular underlying spiral
pattern appears in the infrared, in agreement with Block et al. (1994) and
Thornley (1996). The structure, although very regular, does
sometimes show smaller arms branching off from the main 2--armed structure, and
sometimes a regular 3--armed pattern is revealed. In some cases a 2--armed 
pattern is revealed, but the arms are relatively short. For this reason, we
introduce a scheme for classifying the regularity of spiral arms in the
infrared (to be compared with the arm class of Elmgreen \& Elmegreen 1982). 
We term this classification scheme, the dust--penetrated arm class. Three 
classes are used. Class 1 refers to regular, long arms. Class 2 refers to
regular, but short arms. Class 3 refers to regular arms that have a few 
fragments bifurcating along their length. In the case of class 3, the pattern
still appears much more regular than in the optical, and galaxies referred to
as grand--design based upon their optical appearance, also usually have
many more fragments and bifurcations than those seen in the infrared images
presented in this paper. The dust--penetrated arm class is shown in Table 1
along with comments for some of the galaxies. A digitized sky survey 
inspection of these galaxies has revealed that 6 show optically flocculent 
structure. These galaxies all show Grand Design structure in the 
near--infrared.

We have compared the extent of the spiral structure as revealed in 
near--infrared {\em K} 
band imaging with the rotation curves of these galaxies, in 
order to determine if the spiral structure extends past the radius of 
solid--body rotation. In all 17 galaxies, the spiral structure does extend 
into the differentially rotating part of the disc. This is shown in Table 1
as the fraction of the radius to which grand--design spiral structure extends
divided by 
the radius of turnover in the rotation curve, $r_{gd}/r_{to}$. In all
cases this is significantly greater than 1. It should be noted that it is
possible that in some cases the S/N may not be good enough to trace the
entire spiral structure. In these cases, the ratio $r_{gd}/r_{to}$ is a 
lower limit.

As the ratio $r_{gd}/r_{to}$ is always greater than 1, we proceeded by
testing for
nearby neighbours. Kormendy \& Norman (1979) defined a tidally significant 
neighbour as one within 3--5 galaxy diameters and 1 magnitude ({\em V} band) 
of the spiral galaxy in question. In order to remain consistent with our 
earlier work (Seigar \& James 1998b) we have relaxed this {\em arbitrary} 
requirement to 6 diameters and 3 magnitudes in total brightness. 
We performed a search for near neighbours using 
the Digitized Sky Survey (DSS) and redshifts from the NASA Extragalactic 
Database (NED). We find that 4 of the galaxies have tidally significant 
neighbours using these criteria.

\begin{figure}
\includegraphics{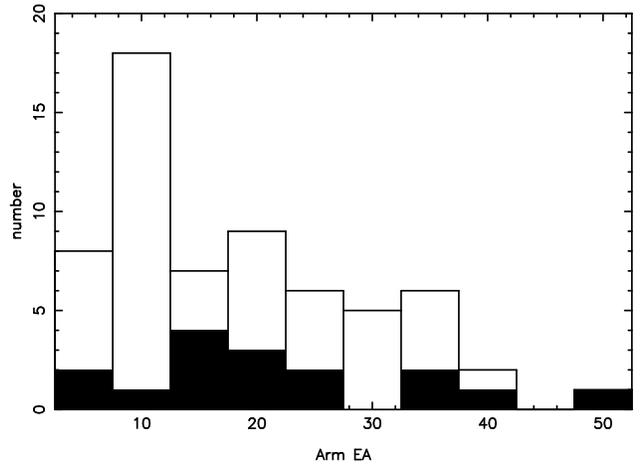}
\vspace*{6.5cm}
\caption{The occurrence of Arm EAs in the discs of a combination of the sample presented in this paper and the sample presented in Seigar \& James (1998b). The numbers shaded in black represent galaxies with companions within six galaxy diameters.}
\end{figure}

\begin{figure}
\includegraphics{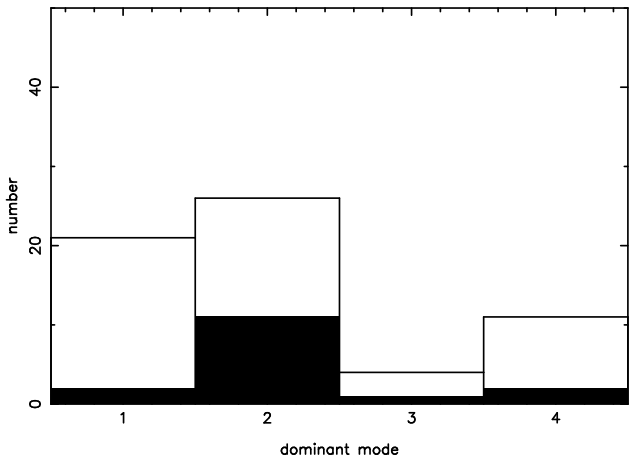}
\vspace*{6.5cm}
\caption{The occurrence of the strongest Fourier modes found in the discs of a combination of the sample presented in this paper and the sample presented in Seigar \& James (1998b). The numbers shaded in black represent galaxies with companions within six galaxy diameters.}
\end{figure}

We also looked at the Hubble classification of these galaxies
(e.g. de Vaucouleurs et al. 1991), in order to see
how many of the galaxies have bars. From this information, 9 of the galaxies
are classified as having bars and 2 are classified as having oval distortions.
We have also investigated the {\em K} band images of the galaxies classified
as non--barred in order to see if the near--infrared imaging reveals bars. The
images of these 6 galaxies are shown in Figure 1. From these images it can be
seen that there is strong evidence for a bar in only one of these galaxies. We 
now describe the images in detail:\\
{\em ESO 576 --G12}: The {\em K} band image of this galaxy shows some evidence
for an asymmetric oval distortion.\\
{\em ESO 602 --G25}: This galaxy shows evidence for a definite oval distortion
in its centre.\\
{\em IC 1330}: This galaxy shows a central bar with arms at its ends.\\
{\em NGC 2722}: The central 3 or 4 contours in this image suggest that there
is some evidence that this galaxy has a weak oval distortion in its centre.\\
{\em UGC 14}: This galaxy has no evidence of a bar in its {\em K} band image.\\
{\em UGC 210}: This galaxy has some elliptical contours (red and green) which
suggest that there is a weak oval distortion present.

The results of the near--neighbour and bar search are listed in Table 1.

Our results suggest that it is difficult to explain the presence of spiral
structure in 5
out of 17 galaxies ($\sim 29$\%) using the
hypothesis of Kormendy \& Norman (1979). This suggests that some other
mechanism may be responsible for driving spiral structure. Indeed along with
the original
hypothesis of a quasi--stationary density wave (Lin \& Shu 1964) it was
proposed that
a density wave is a natural instability of a cold, self--gravitating, stellar
disc.

It should be noted that a difference between this study and the study of
Kormendy \& Norman (1979) is that here we make use of near--infrared images,
whereas Kormendy \& Norman (1979) made use of optical {\em V} band images. 
The near--infrared often reveals underlying Grand--Design spiral structure in
galaxies that exhibit flocculent spiral structure in optical wavebands 
(Thornley 1996), and this is supported by the fact that the galaxies presented
in this paper have regular spiral structure, even though they were not 
selected on the basis of their spiral pattern. In essence, the {\em K} band
studies presented here are therefore touching on aspects of spiral structure
beyond those originally addressed by Kormendy \& Norman (1979).

\begin{figure*}
\includegraphics{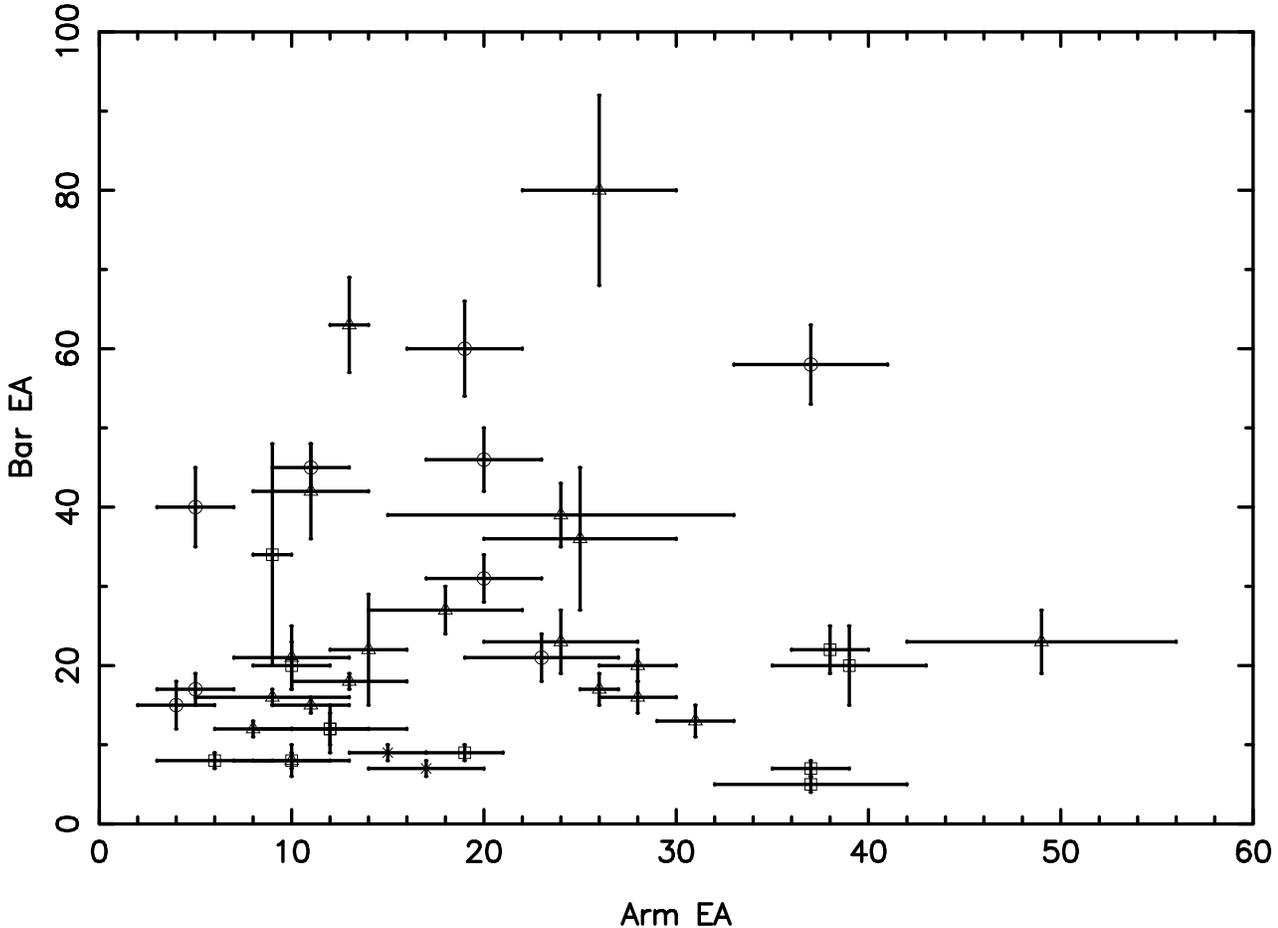}
\vspace*{12cm}
\caption{Arm strength in the inner part of the disc versus bar strength for the barred galaxies in this sample and the sample from Seigar \& James (1998b). Bar strength and arm strength are both measured in terms of EA in degrees. The hollow circles represent galaxies with strong bars in this sample; the crosses represent galaxies with oval distortions in this sample; the hollow triangles represent galaxies with strong bars from the Seigar \& James (1998b) sample; the hollow squares represent galaxies with oval distortions from the Seigar \& James (1998b) sample. Only galaxies classified optically as barred are included.}
\end{figure*}

In the next two sections, we investigate those galaxies with near neighbours
and central bars in more detail, in order to see if either of these properties
are responsible for the driving of spiral structure.

\subsection{Further analysis of galaxies with near neighbours}
\label{sec:nn}

We have measured the strength of arms in this sample of galaxies using
the equivalent angle (EA) method introduced by Seigar \& James (1998a).
The EA is analogous to the quantity equivalent width, which is widely
used in spectroscopy, and is specifically designed to be independent of
noise and resolution effects. Equivalent angle (EA) is defined to be
the angle subtended at the centre of a galaxy by a sector of the 
underlying disc and bulge that emits as much light as the arm/bar
component, within the same radial limits (see Fig. 2 for a 
diagrammatic representation of EA). Thus an arm which emits as much
light as the underlying disc between, say, 0.5 and 1.0 kpc radius
would have an EA of 180$^{\circ}$ within these radial limits. The
same definition is used for bar EA.

Figure 3 shows a histogram of the distribution of arm strengths (in
terms of equivalent angle) in a combination of the samples introduced
here and in Seigar \& James (1998b), making a total of 62 galaxies. The
galaxies with near neighbours are indicated in black in the histogram.
It can be seen that 4 out of 9 galaxies ($\sim 44$\%) with arm EA greater
than 35$^{\circ}$ have near neighbours whereas only 16 out of 62 galaxies
($\sim 26$\%) in the whole sample have companions. A similar and slightly
stronger trend was found by Seigar \& James (1998b), but neither result
is formally statistically significant. However, this result
suggests a relation between arm strength and tidal interactions.

This result is strengthened by Figure 4 which shows that the presence of 
near neighbours enhances the $m=2$ mode, with this being the dominant mode in
11 out of 16 galaxies ($\sim 69$\%)with nearby neighbours, compared with only 
15 out of 46 galaxies ($\sim 33$\%) with no apparent neighbours. 
These statistics are an improvement on
those reported in Seigar \& James (1998b). Thus, it seems that nearby companion
galaxies tend to enhance the strength of the $m=2$ spiral mode.

\subsection{Further analysis of galaxies with bars}
\label{sec:bar}

N--body simulations of disc galaxies as early as Hohl (1971) have shown that
bar modes arise as a natural instability in a cold self--gravitating disc. The
modal theory of spiral structure (Bertin et al. 1989a, b; Bertin \& Lin 1996)
describes bars as a class of self--excited spiral modes. In galaxies with
small bars, the bar can be interpreted as being created by the mode excited
in the outer parts of the disc, when the disc allows for propagation of waves
close to the centre (Bertin \& Lin 1996). However, for strong bars, the
gas response to the bar can still be considered and the modal approach to 
spiral structure takes this into account (Bertin \& Lin 1996). In this case
the original simulations of Sanders \& Huntley (1976) can be taken into 
account. In this section we look at the barred galaxies in this sample, in 
order to see if their spiral modes are bar--driven, and we test the third
part of the Kormendy \& Norman (1979) hypothesis.

Kormendy \& Norman (1979) suggest that the idea of bar
driven spiral modes (Sanders \& Huntley 1976) may be responsible for 
2--armed spiral structure extending past the radius of solid--body 
rotation.
Even if this is the case, Sellwood \& Sparke (1988) state that bar driving
is only important for the most strongly barred spiral galaxies. In addition,
it is thought that bar forcing should not extend far outside the region where
the bar potential is strong.

We have tested models of bar--driven spiral structure by looking for a 
correlation between bar and arm strength. The result of this is shown in
Figure 5, which shows a plot of arm strength in the inner part of the disc
against bar strength. In this plot, we have included the 11 galaxies with bars 
or oval distortions from the sample introduced in this paper, and the 30 
galaxies with bars and oval distortions from the sample introduced in Seigar 
\& James (1998b). We have also split the sample up into those galaxies with 
strong bars and those galaxies with weak oval distortions, by displaying them 
with 
different symbols in the figure. It can be seen that there is no overall 
correlation, and also that there is no correlation for either strongly barred 
galaxies or galaxies with oval distortions. It would therefore appear that 
bars do not strongly affect the strength of spiral arms, even in the central 
regions of discs, and therefore, our conclusions from Seigar \& James (1998b) 
have not changed.

From this, we can  conclude that there is no evidence that bars
are responsible for the driving of
spiral modes in disc galaxies. This, is therefore, evidence therefore against
the part of the hypothesis by Kormendy \& Norman (1979), that bars can drive
spiral structure.

\section{CONCLUSIONS}
\label{sec:conc}

Our $K$--band observations of seventeen inclined spiral galaxies have revealed 
Grand--Design spiral structure in all objects, although six of them
exhibit flocculent structure in the optical. Thus we have introduced the
`dust--penetrated arm class', which indicates the regularity of spiral
structure in the near--infrared. We have used this sample to test the
hypothesis of Kormendy \& Norman (1979) and have found the following:

\begin{itemize}

\item The suggestion by Kormendy \& Norman (1979) that spiral arms may not 
extend
beyond the solid--body part of the rotation curve appears to be rarely, if
ever, the case. 

\item Tidal interactions may play a role in some galaxies, 
but our data also suggests
that many isolated galaxies show strong Grand--Design near--infrared arms.

\item We have found no evidence for driving of spiral structure by bars, and 
support the conclusion of Sellwood \& Sparke (1988) that only the strongest,
relatively rare bars, are likely to have a significant role.

\end{itemize}

Our data have shown that the hypothesis of Kormendy \& Norman (1979)
does definitely not survive the stronger tests posed by near--infrared 
imaging, compared to optical imaging. In this paper we have shown that
some specific mechanism is required to stabilise arm structures, e.g.
density waves (Lin \& Shu 1964, 1966) or modal theories (Bertin et al. 1989a,
b).

\section{acknowledgements}

The United Kingdom Infrared Telescope is operated by the Joint Astronomy 
Centre on behalf of the U.K. Particle Physics and Astronomy Research Council.
This research has made use of the NASA/IPAC Extragalactic Database (NED) which 
is operated by the Jet Propulsion Laboratory, California Institute of
Technology, under contract with the National Aeronautics and Space 
Administration. We thank the anonymours referee for comments which greatly
improved the content of this paper.


\begin{thebibliography}{}

\bibitem{B89a}Bertin G., Lin C.C., Lowe S.A., Thurstans R.P., 1989a, ApJ, 338, 78
\bibitem{B89b}Bertin G., Lin C.C., Lowe S.A., Thurstans R.P., 1989b, ApJ, 338, 104
\bibitem{B91}Bertin G., 1991, in {\em Dynamics of Galaxies and their Molecular Cloud Distributions}, IAU 146, (Kluwer: Dordrecht), p. 93
\bibitem{B93}Bertin G., 1993, PASP, 105, 640
\bibitem{B96}Bertin G., 1996, in {\em New extragalactic perspectives in the New South Africa}, eds Block D.L., Greenberg J.M., (Kluwer: Dordrecht), p. 1
\bibitem{BL96}Bertin G., Lin C.C., 1996, {\em Spiral Structure in Galaxies: A Density Wave Theory}. MIT Press, Cambridge, MA
\bibitem{BW91}Block D.L., Wainscoat R.J., 1991, Nature, 353, 48
\bibitem{BP99}Block D.L., Puerari I., 1999, A\&A, 342, 627
\bibitem{B94}Block D.L., Bertin G., Stockton A., Grosbol P., Moorwood A.F.M., Peletier R.F., 1994, A\&A, 288, 365
\bibitem{dV91}de Vaucouleurs G., de Vaucouleurs A., Corwin H.G. Jr, Buta R.J., Paturel J., Fouqu\'e P., 1991, {\em The Third Reference Catalogue of Bright Galaxies}, Univ. Texas Press, Austin (RC3)
\bibitem{DT94}Donner K.J., Thomasson M., 1994, A\&A, 290, 785
\bibitem{E89}Elmegreen B.G., Elmegreen D.M., Seiden P.E., 1989, ApJ, 343, 602
\bibitem{E82}Elmegreen D.M., Elmegreen B.G., 1982, MNRAS, 201, 1021
\bibitem{E87}Elmegreen D.M., Elmegreen B.G., 1987, ApJ, 314, 3
\bibitem{E00}Eskridge P.B. et al., 2000, AJ, 119, 536
\bibitem{H71}Hohl F., 1971, ApJ, 168, 343
\bibitem{KN79}Kormendy J., Norman C.A., 1979, ApJ, 233, 539
\bibitem{LS64}Lin C.C., Shu F.H., 1964, ApJ, 140, 646
\bibitem{LS66}Lin C.C., Shu F.H., 1966, proc. Natl. Acad. Sci., 55, 229
\bibitem{MFB92}Mathewson D.S., Ford V.L., Buchhorn M., 1992, ApJS, 81, 413
\bibitem{M76a}Miller R.H., 1976a, J. Comp. Phys., 21, 400
\bibitem{M76}Miller R.H., 1976b, ApJ, 207, 408
\bibitem{M78}Miller R.H., 1978, ApJ, 224, 32
\bibitem{SL00a}Salo H., Laurikainen E., 2000a, MNRAS, 319, 377
\bibitem{SL00b}Salo H., Laurikainen E., 2000b, MNRAS, 319, 393
\bibitem{S77}Sanders R.H., 1977, ApJ, 216, 916
\bibitem{SH76}Sanders R.H., Huntley J.M., 1976, ApJ, 209, 53
\bibitem{ST80}Sanders R.H., Tubbs A.D., 1980, ApJ, 235, 803
\bibitem{S02}Seigar M.S., 2002, A\&A, 393, 499
\bibitem{SJ98a}Seigar M.S., James P.A., 1998a, MNRAS, 289, 672
\bibitem{SJ98b}Seigar M.S., James P.A., 1998b, MNRAS, 289, 685
\bibitem{SS88}Sellwood J.A., Sparke L.S., 1988, MNRAS, 231, 25p
\bibitem{S87}Sundelius B., Thomasson M., Valtonen M.J., Byrd G.C., 1987, A\&A, 174, 67
\bibitem{T96}Thornley M.D., 1996, ApJ, 469, L45
\bibitem{TT72}Toomre A., Toomre J., 1972, ApJ, 178, 623
\bibitem{T81}Toomre A., 1981, in {\em The structure and evolution of normal galaxies}, Cambridge University Press, p 111
\bibitem{ZRR93}Zaritsky D., Rix H.--W., Rieke M., 1993, Nature, 364, 313

\end{thebibliography}
\end{document}